\documentclass{article}

\usepackage{microtype}
\usepackage{graphicx}
\usepackage{subcaption} 
\usepackage{booktabs} 

\usepackage{hyperref}

\usepackage[accepted]{icml2025} 

\usepackage{amsmath}
\usepackage{amssymb}
\usepackage{mathtools}
\usepackage{amsthm}
\usepackage{float} 

\usepackage[capitalize,noabbrev]{cleveref} 

\theoremstyle{plain}

\theoremstyle{definition}

\theoremstyle{remark}

\icmltitlerunning{QuantMCP: Grounding LLMs in Verifiable Financial Reality}

\begin{document}

\twocolumn[
\icmltitle{\textit{QuantMCP}: Grounding Large Language Models in Verifiable Financial Reality}

\begin{icmlauthorlist}
\icmlauthor{Yifan Zeng}{sysu} 
\end{icmlauthorlist}

\icmlaffiliation{sysu}{Sun Yat-sen University, Guangzhou, China} 

\icmlcorrespondingauthor{{Yifan Zeng}}{\texttt{yifanzeng0615@foxmail.com}} 

\icmlkeywords{Large Language Models, Financial Technology, Data Accuracy, API Integration, Model Context Protocol, Decision Support, Financial Analysis} 

\vskip 0.3in
]

\printAffiliationsAndNotice{} 

\begin{abstract}
Large Language Models (LLMs) hold immense promise for revolutionizing financial analysis and decision-making, yet their direct application is often hampered by issues of data hallucination and lack of access to real-time, verifiable financial information. This paper introduces QuantMCP, a novel framework designed to rigorously ground LLMs in financial reality. By leveraging the Model Context Protocol (MCP) for standardized and secure tool invocation, QuantMCP enables LLMs to accurately interface with a diverse array of Python-accessible financial data APIs (e.g., Wind, yfinance). Users can interact via natural language to precisely retrieve up-to-date financial data, thereby overcoming LLM's inherent limitations in factual data recall. More critically, once furnished with this verified, structured data, the LLM's analytical capabilities are unlocked, empowering it to perform sophisticated data interpretation, generate insights, and ultimately support more informed financial decision-making processes. QuantMCP provides a robust, extensible, and secure bridge between conversational AI and the complex world of financial data, aiming to enhance both the reliability and the analytical depth of LLM applications in finance.
\end{abstract}

\section{Introduction}
\label{sec:introduction}
The integration of artificial intelligence, particularly Large Language Models (LLMs), into the financial sector heralds a new frontier for data analysis, insight generation, and investment strategy \cite{AIfina}. LLMs possess an extraordinary ability to understand complex natural language queries and synthesize information \cite{histoLens,GPT4}. However, their direct application to tasks requiring precise, up-to-the-minute financial data is fraught with challenges, most notably the propensity for "hallucination" – generating plausible but factually incorrect information \cite{Hallu}. For finance, where accuracy and reliability are non-negotiable, relying on an LLM's internal knowledge for specific market data is untenable and potentially hazardous.

Furthermore, the landscape of financial data is characterized by a multitude of providers, each with unique APIs, data structures, and access protocols \cite{FinAPI}. Bridging the gap between an LLM's conversational interface and this fragmented, programmatic world requires more than ad-hoc solutions; it demands a structured, secure, and scalable approach. The emerging Model Context Protocol (MCP) offers a promising standard for enabling LLMs to discover and interact with external tools and data sources in a controlled manner \cite{MCP}.

This paper introduces \textbf{QuantMCP}, a framework engineered to address these critical challenges. QuantMCP's primary objective is to ground LLMs in verifiable financial reality, thereby transforming them from potentially unreliable information sources into powerful engines for accurate data retrieval and sophisticated financial analysis. It achieves this by: (1) Utilizing an LLM's natural language understanding to interpret user requests for financial information. (2) Employing the MCP to enable the LLM to select, parameterize, and invoke appropriate "tools," which are essentially wrappers around functions from diverse Python-accessible financial data APIs (e.g., Wind, yfinance, Tushare). (3) Ensuring that all financial data presented or analyzed by the LLM is sourced directly and in real-time (or as per API capability) from authoritative providers, thus mitigating hallucination and guaranteeing data integrity. 


Once QuantMCP furnishes the LLM with accurate, structured financial data, the LLM's role transcends mere data fetching. It can then leverage its inherent analytical prowess—such as pattern recognition, summarization, comparative analysis, and even hypothesis generation—to empower deeper financial insights and support complex decision-making processes. For instance, an LLM, having accurately retrieved historical stock prices and trading volumes via QuantMCP, could then be prompted to identify trends, calculate volatility, or compare performance against benchmarks, tasks it could not reliably perform without grounded data.

The key contributions of QuantMCP can be summarized as:
\begin{itemize}
    \item \textbf{Enhancing Data Reliability for LLMs in Finance:} Systematically combating LLM hallucination by ensuring all financial data is retrieved from verified sources via programmatic API calls.
    \item \textbf{Democratizing Access to Sophisticated Financial Tooling:} Lowering the technical barrier for interacting with financial data and leveraging advanced AI analytics, making these capabilities accessible to a broader range of users.
    \item \textbf{Unlocking LLM-Powered Financial Analysis on Grounded Data:} Enabling LLMs to move beyond simple Q\&A to perform meaningful analytical tasks on accurately retrieved financial information, thus augmenting the capabilities of financial professionals.

\end{itemize}

This paper will elaborate on the architecture of QuantMCP, discuss the design of its core components including the MCP server, present a prototype implementation, and illustrate its functionality through practical case studies. We will conclude by discussing the profound implications of QuantMCP for transforming financial workflows, and promising directions for future development in building trustworthy and potent AI-driven financial systems.
\begin{figure*}[t]
  \centering
  \includegraphics[width=\linewidth]{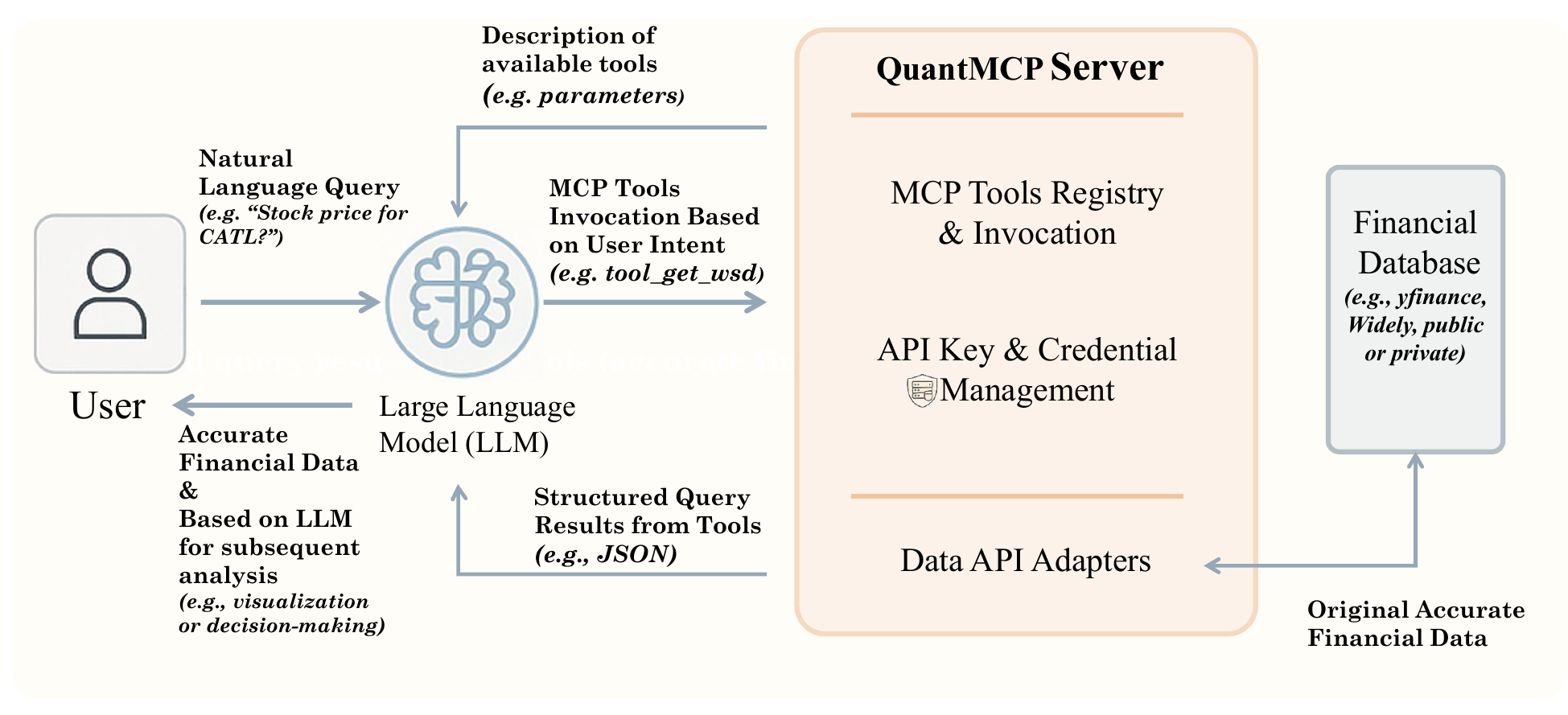} 
  \caption{Architecture of the QuantMCP Framework. This diagram shows the interaction flow from user natural language query, through the LLM and QuantMCP Server, to the financial database, and back to the LLM for analysis and response generation.}
  \label{fig:quantmcp_architecture}
\end{figure*}

\section{Background}
\label{sec:background}
The QuantMCP framework is predicated on the synergistic advancements in Large Language Models (LLMs), the standardization efforts for model-tool interactions as exemplified by the Model Context Protocol (MCP), and the dynamic Python ecosystem for accessing financial data.

Large Language Models, with prominent examples based on the Transformer architecture (e.g., GPT series, Llama, Claude) \cite{GPT4}, have evolved beyond their initial text generation capabilities to become powerful engines for natural language understanding and complex reasoning \cite{CoT}. A particularly transformative development is their ability to function as intelligent agents interacting with external software tools and APIs. Through sophisticated prompting techniques and agentic frameworks, LLMs can now interpret complex user requests, decompose them into actionable steps, identify and parameterize appropriate tools (function calls or API endpoints) from a given repertoire, and process returned information to synthesize responses or drive further actions. However, their direct application in domains requiring high factual accuracy, such as finance, is severely challenged by the phenomenon of "hallucination"—the confident generation of incorrect or fabricated information \cite{Hallu}. This inherent limitation underscores the critical need for robust mechanisms to ground LLM operations in verifiable external data sources, especially when dealing with dynamic and precise financial data. QuantMCP addresses this by positioning the LLM as an intelligent orchestrator that directs calls to authoritative data providers, rather than being a primary source of financial facts itself. Importantly, once furnished with accurate, externally sourced data, LLMs can then leverage their advanced analytical capabilities to perform tasks such as trend summarization, anomaly detection, and even generating explanatory narratives for financial patterns \cite{LLMFina,LLMFina1}, thereby unlocking significant value.

The increasing integration of LLMs with external systems highlights the urgent need for standardized, secure, and discoverable methods for model-tool interaction. The Model Context Protocol (MCP) \cite{MCP} is an emerging open standard designed to meet this demand. MCP aims to establish a common "language" and interaction paradigm between LLMs (or LLM-powered agents) and "MCP Servers," which expose a suite of "tools" representing callable functions or services. Core to MCP's philosophy are the principles of tool discovery, where servers declare their capabilities via a manifest, and standardized invocation, which simplifies integration and promotes interoperability. Furthermore, MCP inherently supports context management, allowing relevant information (such as conversation history or user state) to be passed to tools for more nuanced operations. Another advantage of the MCP architecture, crucial for domains like finance, is that it provides a natural control point for implementing robust security measures. This robust, standardized approach can be generalized across various financial APIs, ensuring both reliability and scalability in how LLMs access and utilize external financial data tools.

QuantMCP can provide a standardized framework for interfacing multiple distinct LLMs with a wide array of data sources, ranging from comprehensive commercial providers like Wind Information and Bloomberg, to brokerage APIs offering real-time trading data, and numerous open or freemium data sources such as yfinance, Tushare, and Alpha Vantage.

\section{The QuantMCP Framework}
\label{sec:quantmcp_framework}
QuantMCP is architected as a robust and extensible system that seamlessly integrates diverse Large Language Models (LLMs) with a wide array of financial data APIs via the Model Context Protocol (MCP). Its design prioritizes data accuracy, secure API interaction, and the empowerment of LLMs to perform complex downstream analytical tasks, moving beyond simple data retrieval. \cref{fig:quantmcp_architecture} illustrates its conceptual architecture.


\subsection{Detailed Description}

\textbf{Natural Language Interface (NLI) Layer (LLM-Powered):} This is the primary interaction point for the user. An LLM (e.g., GPT-4, Gemini 2.5, DeepSeek-V3) serves as the NLI, responsible for understanding the user's financial data query expressed in natural language. Its tasks include intent recognition (what data is needed), entity extraction (e.g., specific stock tickers, date ranges, financial indicators), and query disambiguation. Crucially, the LLM does not attempt to answer factual data queries from its parametric memory; instead, it formulates a plan to retrieve this information using available tools.

\textbf{MCP Server for Financial Tools:} At the core of QuantMCP, the MCP Server acts as a standardized gateway between the LLM and financial data retrieval functionalities. It hosts a collection of "tools," each representing a specific function for accessing data from one or more financial APIs. The server exposes these tools to the LLM via an MCP-compliant manifest, detailing each tool's purpose, required parameters (with types and descriptions), and expected output format. QuantMCP generalizes this concept to any Python-accessible API. The MCP Server is also responsible for securely managing API credentials.

When a tool (via an adapter, implicitly managed by the MCP server or explicitly defined for different APIs) retrieves data from an external API, it is crucial that this data is returned in a structured, LLM-consumable format, typically JSON. This structured data, now grounded in a verifiable external source, is then passed back to the LLM. This step is fundamental to overcoming LLM hallucination, as the LLM's subsequent analysis or response generation is based on this verifiably accurate dataset.

\textbf{LLM for Analysis, Synthesis, and Decision Support:} Once the LLM receives the structured, grounded data from the MCP Server, its role shifts from query understanding to data utilization. It can now perform a wide range of analytical tasks, such as summarizing retrieved data, identifying trends or anomalies, performing calculations or comparisons, generating textual explanations based on the data, and providing contextually relevant information that can support human investment decisions or feed into more complex algorithmic trading strategies. This is where QuantMCP truly empowers the LLM, transforming it into a sophisticated analytical assistant that operates on a foundation of trustworthy information.

A typical workflow based on QuantMCP can be summarized as follows: A user poses a natural language query. The LLM (NLI Layer) parses this query and determines the need to invoke one or more financial tools. It then interacts with the MCP Server, providing the appropriate tool name(s) and extracted parameters. The MCP Server invokes the tool(s), which utilize external financial APIs to retrieve data, structure it, and then return it to the LLM. Finally, the LLM processes this grounded data to generate an analytical response, a summary, or further questions to the user, potentially leading to iterative data retrieval and analysis cycles. This entire process is designed to be transparent to the end-user, who experiences it as a seamless natural language conversation leading to accurate, data-backed financial insights.


\subsection{Ensuring Data Accuracy and Mitigating Hallucination}
A cornerstone of the QuantMCP framework is its commitment to data accuracy. Unlike approaches where an LLM might attempt to recall financial data from its training corpus (which is static and prone to inaccuracies or outdatedness), QuantMCP strictly mandates that all specific financial data points are fetched from live, authoritative APIs at the time of the query (or from the most recent available data from the API). The LLM's role is to understand what data is needed and how to ask for it via the MCP tools, not to provide the data itself from its internal knowledge. This "grounding" in external, verifiable sources is the primary mechanism by which QuantMCP mitigates the risk of LLM hallucination in the context of financial data retrieval. The structured nature of API responses, further standardized into JSON or similar formats by the API adapters, also ensures that the LLM receives data in a clear, unambiguous format, reducing the chance of misinterpretation during subsequent analytical tasks.

\subsection{Security Considerations within MCP}
Handling financial data and API credentials necessitates robust security measures. The MCP Server within QuantMCP is strategically positioned to manage these concerns. API keys and authentication tokens for various financial data services are stored and managed securely on the server-side, within the scope of the API adapters or the MCP server itself. They are never exposed to the LLM or the client-side application. When the LLM requests a tool invocation, the MCP server authenticates the request (potentially based on user session or LLM identity) and then uses the appropriate stored credentials to interact with the backend financial API. This centralized credential management significantly reduces the attack surface.

\subsection{User-Friendly Financial Data Access}

A significant innovation of the QuantMCP framework lies in its ability to democratize access to complex financial data through an intuitive natural language interface. Traditionally, retrieving specific financial datasets requires users to learn and navigate intricate API documentation, understand specific parameter names, data codes (e.g., ticker symbols, indicator codes), and often write programmatic scripts. This presents a steep learning curve and effectively excludes a large population of potential users who may possess financial acumen but lack advanced programming skills.

QuantMCP fundamentally simplifies this interaction paradigm. By leveraging the advanced natural language understanding (NLU) capabilities of LLMs, users can express their data needs in plain, conversational language. For example, instead of needing to know the exact API function and parameters for historical stock prices, a user can simply ask, "\textit{What were the daily high, low, and closing prices for Microsoft and Amazon from January 1st, 2023, to December 31st, 2023?}". The NLI layer within QuantMCP (powered by an LLM) is responsible for parsing this request, identifying the key entities (Microsoft, Amazon, date range, desired price types), and mapping them to the corresponding parameters of the appropriate MCP tool.

This natural language front-end significantly lowers the technical barrier for sophisticated data retrieval, making it accessible to a broader audience, including non-programmers, and enhances efficiency for all users by streamlining data acquisition. Furthermore, its conversational nature improves data discoverability and, by simplifying access to accurate financial data, QuantMCP contributes to the democratization of financial insights, empowering wider data-driven exploration and fostering greater financial literacy. The framework's core strength in translating complex, often implicit, natural language needs into precise, structured API calls is pivotal to this user-centric approach, ensuring both accessibility and the retrieval of relevant, accurate information that faithfully represents user intent.

\section{Implementation and Case Studies}
\label{sec:implementation_case_studies}
To demonstrate the practical viability and effectiveness of the QuantMCP framework, we developed a prototypical implementation and conducted several case studies mirroring common financial data retrieval and preliminary analysis tasks. Our prototype prioritizes modularity to allow for the integration of various LLMs and financial data APIs.

\subsection{Prototypical Implementation Details}
Our QuantMCP prototype consists of the core components described in \cref{sec:quantmcp_framework}:

\begin{figure}[t]
\centering
\includegraphics[width=\columnwidth]{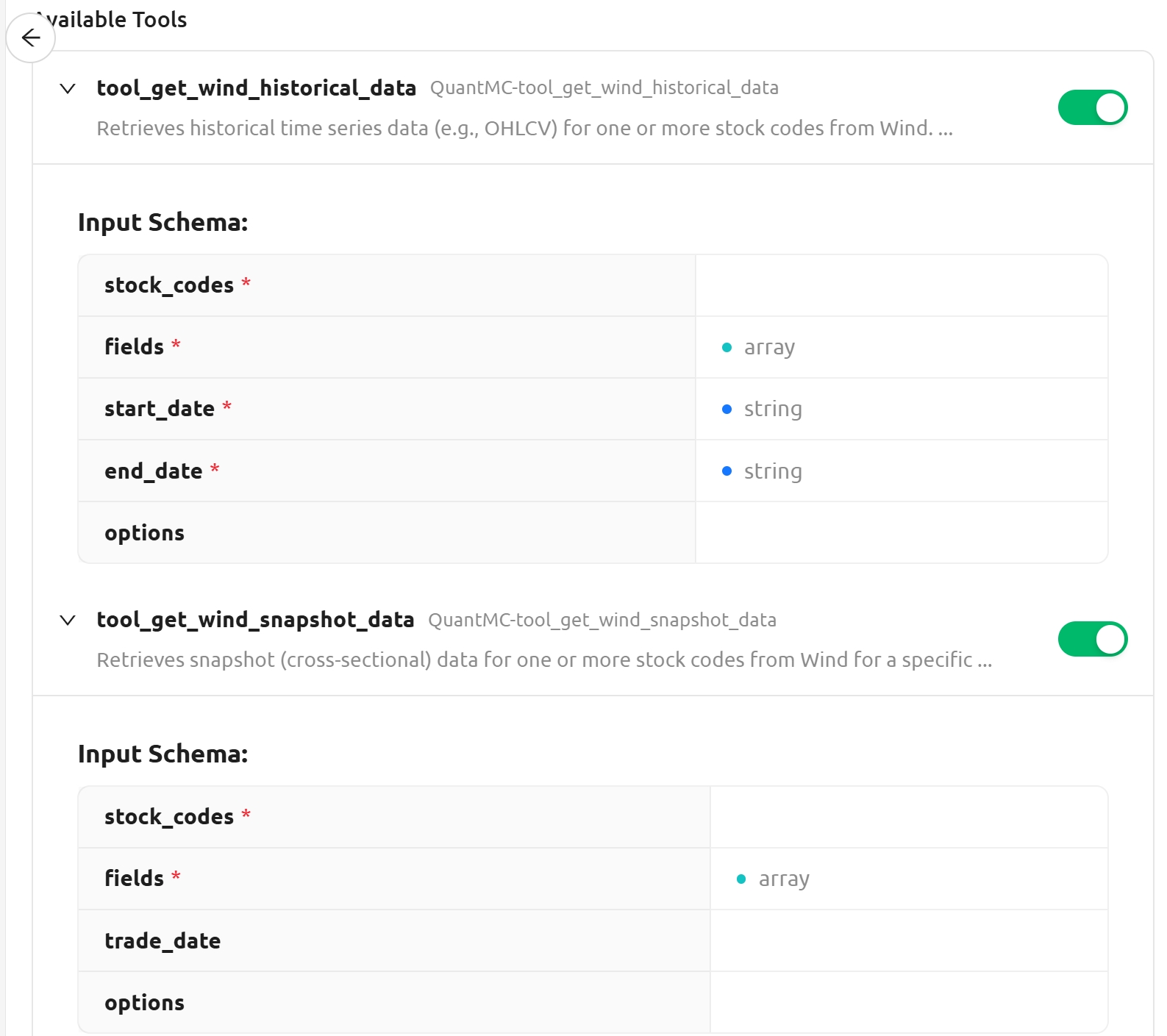}
\caption{Part of tools provided by Wind API's QuantMCP Server, including explanations and parameter requirements.}
\label{fig:Tool}
\end{figure}

\begin{itemize}
    \item \textbf{Natural Language Interface (NLI) Layer:} We utilized a state-of-the-art Large Language Model, specifically DeepSeek-V3-0324 (671B) \cite{DeepSeek} (as an example, interchangeable with models like GPT-4, Gemini 2.5 Pro or Claude 3.7 Sonnet), known for its strong instruction-following and reasoning capabilities. The LLM interacts with the user and the MCP server. We use Cherry Studio\footnote{\url{https://github.com/CherryHQ/cherry-studio}} to integrate LLM with MCP Server.
    \item \textbf{MCP Server:} We implemented an MCP server. This server hosts a set of financial data tools. Each tool is defined with a clear name, description, input parameters (with types), and expected output format, all exposed via an MCP manifest.

    \item \textbf{Financial Data API:} Data provided by Wind Information Co., Ltd. The decision to utilize the \textbf{Wind API}\footnote{\url{http://www.wind.com.cn/}} as a core data source in our prototypical QuantMCP implementation stems from its position as a leading financial information provider. Wind offers {data depth, breadth, and authoritativeness}, encompassing equities, bonds, funds, derivatives, macroeconomic indicators, and industry-specific data, which are crucial for sophisticated financial research and professional investment decision-making. The \textit{\textit{WindPy} } library provides {powerful and extensive API functionalities} that allow for complex data queries and access to pre-calculated analytics directly from the Wind terminal. This tight integration ensures data timeliness and leverages the terminal's robust computational capabilities. While \textit{WindPy}  necessitates a local Wind terminal installation and login, demonstrating its integration within QuantMCP highlights the framework's capacity to empower LLMs with access to {professional-grade, high-fidelity financial intelligence}.

\end{itemize}

Ensuring the security of external API calls, such as those to Alpha Vantage or Tushare, is a foundational design principle within the QuantMCP framework. API keys are {strictly confined to secure server-side storage and centralized management within the QuantMCP server}, rendering them inaccessible to the LLM itself or to non-administrative end-users. This architecture establishes a security barrier, effectively mitigating the risks associated with direct API key exposure and uncontrolled API invocations.


\subsection{Illustrative Workflow: Querying and Analyzing Stock Data with QuantMCP}
\label{ssec:workflow_example_windpy_cn_detailed_paragraph}

\begin{figure}[t]
\centering
\includegraphics[width=\columnwidth]{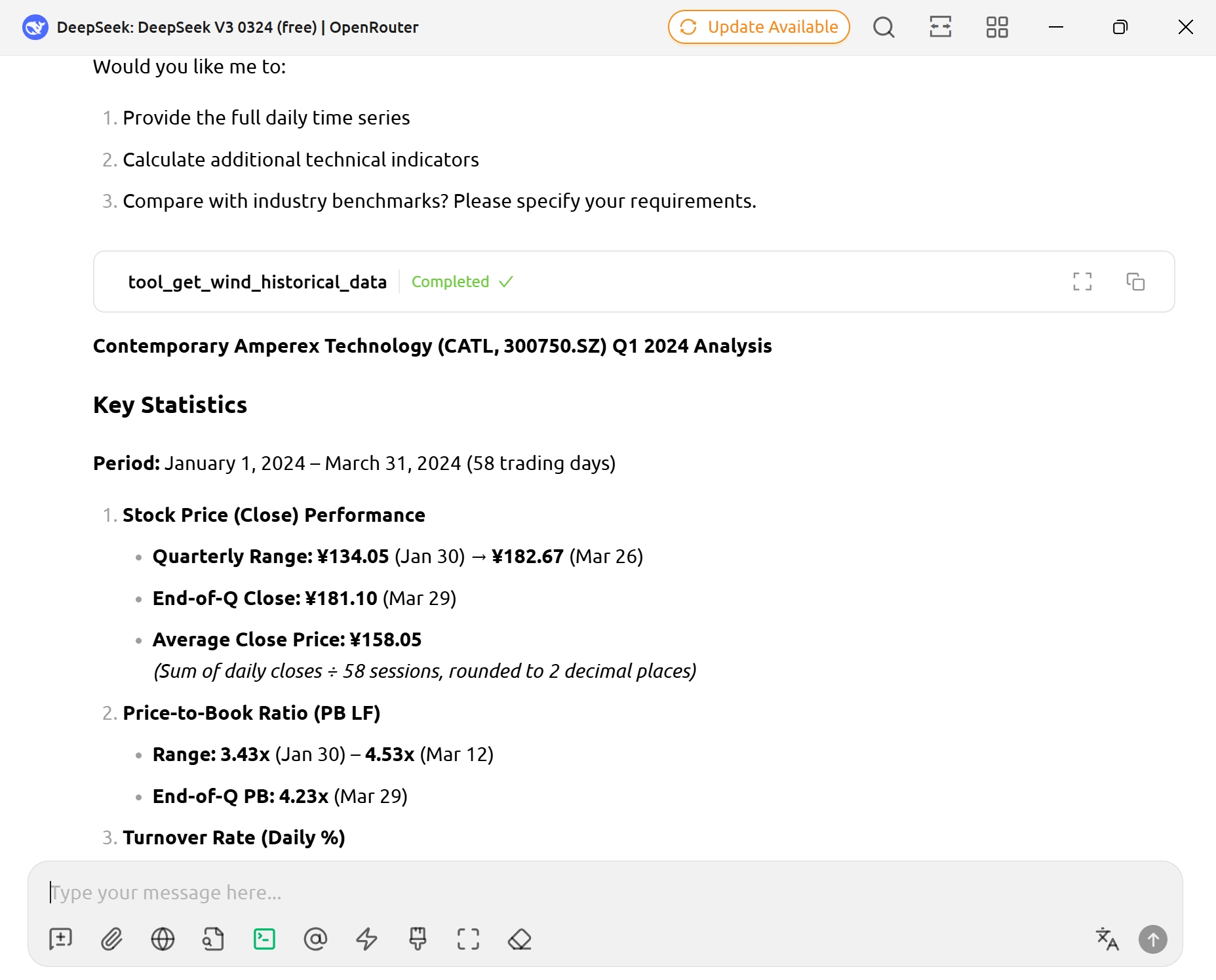}
\caption{Representation of the response generated by the LLM (DeepSeek-V3-0324) after processing the user query through the QuantMCP workflow, displaying retrieved data and calculated averages. }
\label{fig:model_response_wind_placeholder_italic}
\end{figure}

\begin{figure}[t]
\centering
\includegraphics[width=\columnwidth]{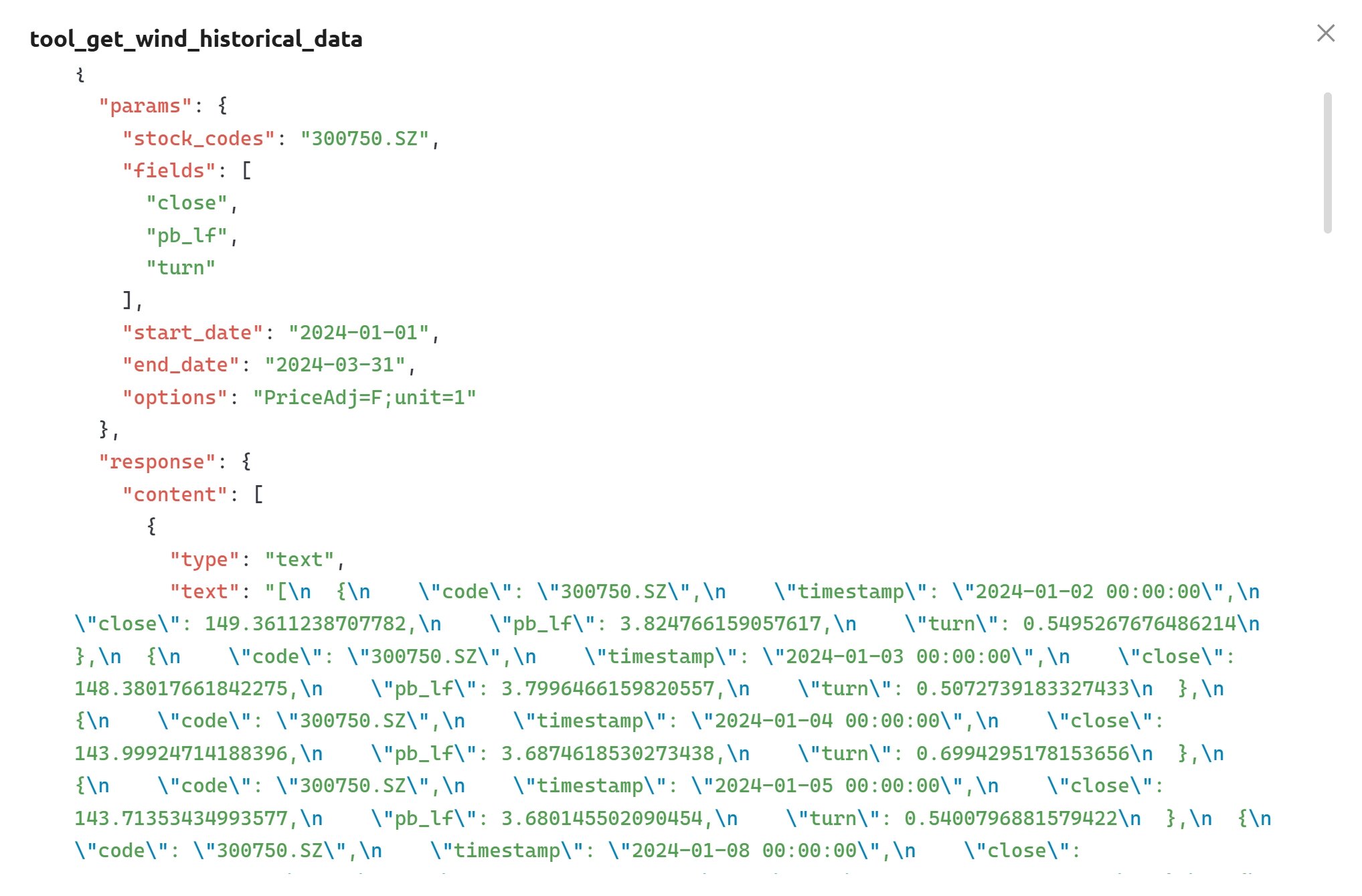}
\caption{Part of the JSON-formatted returned results obtained by the LLM's automatic MCP tool invocation are strictly based on the precise data results returned by the API. The LLM accurately acquires this structured data as the foundation for overcoming hallucinations. }
\label{fig:JSON}
\end{figure}

\begin{figure}[h]
\centering
\includegraphics[width=\columnwidth]{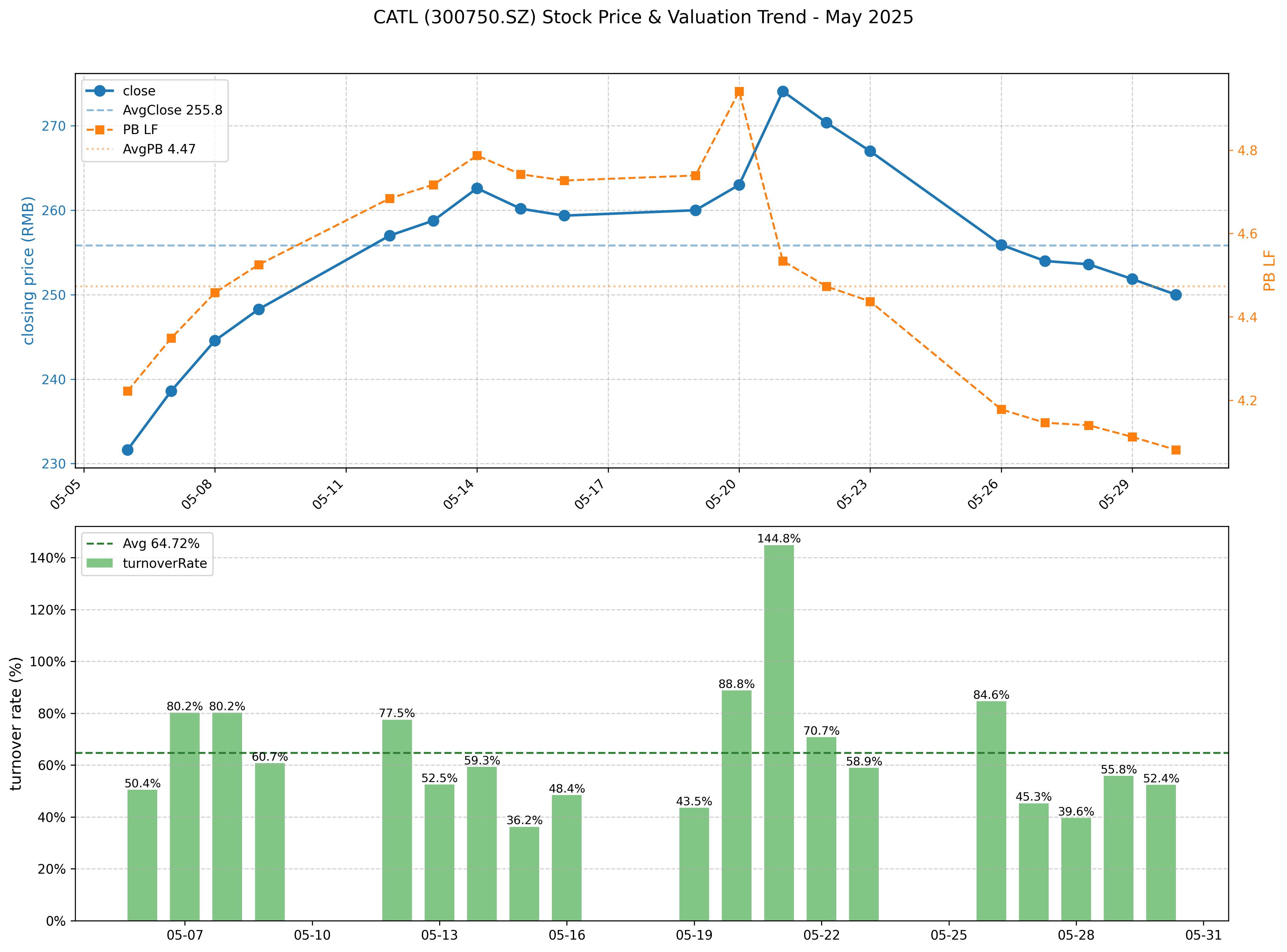}
\caption{Leveraging the LLM's coding capabilities, we conduct further visualization analysis of the precise financial data from QuantMCP integrated with the LLM. Here, we focus on CATL's closing price, P/B ratio, and other key metrics for May 2025.}
\label{fig:Vis}
\end{figure}

To further illustrate the operational flow, consider a user prompt: \textit{Query the daily closing price, price-to-book ratio (PB LF), and turnover rate (percentage) of Contemporary Amperex Technology (300750.SZ) from the beginning to the end of the first quarter of 2024. Please tell me its average closing price and average turnover rate during this quarter.}
Upon receiving this, the LLM (e.g., DeepSeek-V3-0324) first parses the query, identifying the stock code (\textit{300750.SZ}), the required historical daily indicators (\textit{close}, \textit{pb\_lf}, \textit{turn}), and the precise date range (Q1 2024, translated to \textit{"2024-01-01"} to \textit{"2024-03-31"}). It also recognizes the analytical tasks: calculating average closing price and average turnover rate. The LLM then determines that the \textit{tool\_get\_wind\_historical\_data} MCP tool is appropriate and formulates a structured JSON call, specifying the stock code, fields (including \textit{"timestamp"} for date context), the calculated date range, and relevant Wind options like \textit{"PriceAdj=F;Fill=Previous"}. This MCP request is dispatched to the QuantMCP server, which, after ensuring an active \textit{WindPy} connection, invokes the underlying Python function. This function, in turn, executes the \textit{w.wsd} command to fetch data directly from the Wind terminal. The raw \textit{WindData} object is then parsed and standardized into a JSON-compatible list of dictionaries by the server (e.g., \textit{[{"code": "300750.SZ", "timestamp": "2024-01-02 15:00:00", "close": 180.50, ...}, ...]}) and returned to the LLM. Critically, with this verified, structured data from Wind, the LLM proceeds to perform the requested calculations—arithmetic means for \textit{close} and \textit{turn} values. Finally, the LLM synthesizes all retrieved and computed information into a comprehensive natural language response for the user, presenting both the raw data insights (e.g., sample daily figures) and the derived analytical results (the calculated averages), thereby demonstrating QuantMCP's end-to-end capability to ground LLM analysis in accurate, externally sourced financial data. The model's final response, illustrated in Figure \ref{fig:model_response_wind_placeholder_italic}, would clearly present these findings.

We can conduct further experiments to obtain precise financial data, then leverage the LLM's coding capabilities for visualization. By utilizing methods similar to Jupyter MCP, we can automatically generate and execute visualization code to produce clear, accurate, visually appealing, and fully automated visualization results. This will better present the precise data in a more readable format. Here, we use DeepSeek-V3-0324 and tools provided by the QuantMCP framework to query CATL's closing price, P/B ratio, and other data for May 2025, followed by further visualization (as shown in \cref{fig:Vis}).

\section{Discussion}
\label{sec:discussion}

The QuantMCP framework, as presented, offers a significant advancement in how Large Language Models can be effectively and reliably integrated into the financial domain, particularly for accessing and analyzing dynamic market data. Its core strength lies in rigorously grounding LLM interactions in verifiable data retrieved directly from authoritative APIs, thereby mitigating the critical issue of LLM hallucination which is particularly detrimental in finance. By leveraging the Model Context Protocol (MCP), QuantMCP introduces a standardized, secure, and extensible mechanism for LLMs to discover and utilize a diverse array of financial data tools built upon Python APIs like WindPy or Alpha Vantage. This protocol-driven approach moves beyond ad-hoc API integrations, paving the way for more robust and interoperable AI-powered financial systems. The natural language interface provided by the LLM dramatically lowers the technical barrier to entry for sophisticated financial data retrieval and initial analysis, empowering a broader range of users, from financial analysts seeking to accelerate their workflows to non-programmers aiming to make data-informed decisions.

The profound implications of QuantMCP extend beyond mere data accessibility. By furnishing LLMs with accurate, timely, and structured financial data, the framework unlocks their latent analytical capabilities. Once grounded, LLMs can perform a variety of value-added tasks, such as identifying trends, calculating key performance indicators, summarizing complex datasets, generating comparative analyses, and even flagging potential anomalies or correlations that might warrant human attention. This synergy transforms the LLM from a simple conversational agent or data fetcher into a powerful analytical assistant, capable of augmenting human expertise and potentially streamlining parts of the investment research and decision-making pipeline. For instance, an LLM could autonomously gather relevant historical performance, fundamental data, and market sentiment indicators for a portfolio of stocks via QuantMCP tools, and then synthesize this information into a concise daily briefing for a fund manager, a task that would traditionally require significant manual effort. This shift towards LLM-driven, data-grounded analysis could lead to increased efficiency, novel insights derived from complex data interactions, and a more democratized access to sophisticated analytical tooling within the financial industry.
\nocite{langley00}

\bibliography{example_paper}
\bibliographystyle{icml2025}




\end{document}